\documentclass[nofootinbib,aps,11pt]{revtex4}
\usepackage{epsfig}
\usepackage{amsmath}
\usepackage{bm}
\usepackage{times}
\usepackage{graphicx}
\usepackage{color}

\def\nn{\nonumber}
\def\Eq#1{Eq.~(\ref{#1})}

\def\stilde{\widetilde}

\def\lagr{{\cal L}}

\begin{document}

\preprint{IFIC/09-XX}

\title{\Large Gauge Mediated SUSY Breaking via Seesaw}
\author{Pavel Fileviez P\'erez$^{1}$}
\author{Hoernisa Iminniyaz$^{2}$}
\author{Germ\'an Rodrigo$^{3}$}
\author{Sogee Spinner$^{1}$ \\}
\affiliation{\\  \\ $^{1}$University of Wisconsin-Madison, Physics Department\\
1150 University Avenue, Madison, WI 53706, United States of America.\\
$^{2}$Center for High Energy Physics, Peking University, Peking, \\ 
and Physics Department, University of Xinjiang, Urumqi, China. \\
$^{3}$ 
Instituto de F\'{\i}sica Corpuscular,
UVEG-Consejo Superior de Investigaciones Cient\'ificas,
Apartado de Correos 22085, E-46071 Valencia, Spain}
\date{\today}
\begin{abstract}
We present a simple scenario for gauge mediated supersymmetry breaking where the messengers 
are also the fields that generate neutrino masses. 
We show that the simplest such scenario corresponds to the case where neutrino 
masses are generated through the Type I and Type III seesaw mechanisms.  
The entire supersymmetric spectrum and Higgs masses are calculable from 
only four input parameters.  Since the electroweak symmetry is broken through a doubly 
radiative mechanism, meaning a nearly zero $B$-term at the messenger scale which 
runs down to acceptable values,  one obtains quite a constrained spectrum for the 
supersymmetric particles whose properties we describe. We refer to this mechanism as ``$\nu$-GMSB".
\end{abstract}
\maketitle

\section{Introduction}
The Minimal Supersymmetric Standard Model (MSSM) is one of the most appealing 
extensions of the Standard Model with a mechanism to protect the Higgs mass 
from radiative corrections, realizable high scale gauge coupling unification~\cite{Unif1,Unif2,Unif3,Unif4}, 
and a candidate for the cold dark matter of the 
Universe, even when the so-called discrete $R$-parity symmetry is broken. Furthermore,
the mechanism of electroweak baryogenesis can be employed to explain 
the matter-antimatter asymmetry in the Universe and one has the appealing mechanism 
for radiative electroweak symmetry breaking (EWSB)~\cite{REWSB}.

One of the open issues in the MSSM is the origin of  supersymmetry (SUSY) 
breaking (see Ref.~\cite{Dine-review} for a review on supersymmetry breaking).
Gauge Mediation~\cite{GMSB} is one of the most appealing mechanisms to address this
issue.  Superpartner masses are predicted assuming the existence of a SUSY breaking hidden sector.  
This breaking is then transmitted to the MSSM sector through gauge interactions via 
messenger fields.  This process generates masses for all superpartner 
masses and avoids the so-called flavor problem in SUSY theories since
mixings between the sfermions of different families are not generated. 

In this paper we present a simple scenario for gauge mediated supersymmetry breaking
where the messengers are also the fields that generate neutrino masses.
We refer to this scenario as ``\textrm{$\nu$-GMSB}".  We build up to this
scenario by discussing previous implementations of gauge mediation through the already existing
particle content of the simplest $SU(5)$ supersymmetric grand unified theories (GUT). Since it is expected
that in such models we will also generate neutrino masses in a consistent way, we then consider
seesaw extensions of $SU(5)$.  The so called seesaw fields can also mediate SUSY breaking 
and since the seesaw scale, $M_{\textrm{Seesaw}}  \lesssim 10^{11-14}$ GeV, is much 
smaller than the GUT scale, $M_{\textrm{GUT}} \approx 10^{16-17}$ GeV, the seesaw contributions will
dominate the SUSY breaking masses.  This idea was first discussed in Refs.~\cite{Farzan,Rossi,Mohapatra}. 

We investigate this hypothesis discussing all possible scenarios 
for gauge mediation in the context of  $SU(5)$ theories and find that
neutrino mass generation through both the Type I and Type III seesaw mechanisms provides 
the simplest framework for gauge mediation via seesaw fields.  We then pursue 
this idea in detail, finding that the spectrum depends on four parameters and that 
while the bilinear Higgs term is very small at the messenger scale, it can run to 
acceptable values for electroweak symmetry breaking (EWSB) at the SUSY scale.  
We obtain a constrained SUSY spectrum whose phenomenological aspects are then discussed.

In Section II we discuss the different implementation of the gauge mediation mechanism 
for supersymmetry breaking in the context of $SU(5)$ grand unified theories. 
In Section III we discuss the predictions for superpartner masses in the case where 
the messengers are the fields responsible for the Type III seesaw mechanism. 
In Section IV we discuss the constraints from gauge coupling unification 
and proton decay, while in Section V we summarize our findings. 
\section{Supersymmetric SU(5) Unification and Gauge Mediated SUSY Breaking}
In the minimal supersymmetric $SU(5)$~\cite{SUSY-SU(5)} the MSSM matter fields of one family are unified in 
${\bf \hat{\bar{5}}}= (\hat{D}^C,\hat{L})$, and ${\bf \hat{10}}= (\hat{U}^C,\hat{Q},\hat{E}^C)$, while the 
Higgs sector is composed of  $\mathbf{\hat 5_H}=(\hat{T}, \hat{H})$, $\mathbf{\hat{\bar5}_H}=(\hat{\overline{T}},\hat{\overline{H}})$, 
and $\mathbf{\hat{24}_H}=(\hat{\Sigma}_8,\hat{\Sigma}_3, \hat{\Sigma}_{(3,2)}, \hat{\Sigma}_{(\bar{3}, 2)}, \hat{\Sigma}_{24})
=(8,1,0)\bigoplus(1,3,0) \bigoplus (3,2,-5/6)\bigoplus (\overline{3},2,5/6)\bigoplus (1,1,0)$. 
In this model the Yukawa superpotential for charged fermions reads as:
\begin{eqnarray}
{\cal{W}}_{1} = {\hat{10}} \  Y_1 \ \hat{\overline{{5}}} \ \hat{\overline{5}}_H
\ +   {\hat{10}} \  Y_2 \ {\hat{10}}  \  {\hat{5}_H}  \ + \  {\cal O}(24_H/M_{Pl}).
\label{W1}
\end{eqnarray}
Here we assume the existence of higher-dimensional operators for consistent
fermion masses. The relevant interactions for breaking $SU(5)$ to the SM gauge group
are:
\begin{eqnarray}
{\cal W}_{2} = m_{\Sigma} \ \text{Tr} \ \hat{24}_H^2 \ + \
{\lambda_{\Sigma}} \  \text{Tr} \ \hat{24}_H^3 \ + \  {\cal O}(24_H/M_{Pl}),
\end{eqnarray}
and the interactions between the different Higgs chiral superfields are:
\begin{eqnarray}
{\cal W}_{3} &=& m_{H} \ \hat{\overline{5}}_H  \hat{5}_H \ + \
\lambda_H \ \hat{\overline{5}}_H  \hat{24}_H \hat{5}_H \ + \  {\cal O}(24_H/M_{Pl}).
\label{HMixing}
\end{eqnarray}
See Refs.~\cite{review,BPG1,BPG2} and references therein for the status of this model. 

At first sight, the most appealing way to proceed is to break both the GUT symmetry and SUSY
with the same field, ${\bf \hat{24}_H}$~\cite{Agashe,Borut-Goran} assuming,
$\left<\hat{24}_H\right>= v_{24} \ + \ \theta^2 \ F_{24}$.  This would lead to
contributions to the soft terms from the following
four sets of would-be messengers:
the Higgses---$H$ and $\overline{H}$~\cite{Dvali-Shifman}; the $SU(3)$ color triplets---$T$ and
$\overline{T}$; the components of
the ${\bf \hat{24}_H}$---$\Sigma_3$ and $\Sigma_8$; and the $SU(5)$ heavy gauge bosons---
$X$ and $Y$~\cite{Dermisek-Kim}.  The largest contribution by far is the first one, via the Higgses, since they are the lightest of these fields.  SUSY breaking in this case
is transmitted once we generate the term $H \overline{H} F_{24}$ using the 
scalar interactions from Eq.~(\ref{HMixing}). Unfortunately, this possibility is
ruled out due to negative leading order
contributions to the sfermion masses, which produce a tachyonic stop~\cite{Dine}.
While the coupling of the Higgses to the ${\bf \hat{24}_H}$ must exist in order to achieve
double-triplet splitting, it is important to note that the $H \overline{H} F_{24}$ term can be eliminated 
by invoking extra fine-tuning on top of the fine-tuning needed for doublet-triplet splitting
\footnote{One could avoid this problem adding several 24 representations~\cite{Melfo}}.
We find this possibility unappealing and will not discuss it further.
 
Since the fields present in the theory cannot be used to transmit SUSY breaking, the simplest approach
is to introduce a new singlet field, $\hat{S}$, which couples both to the $SU(5)$ visible 
sector and to the hidden sector.  We assume this singlet does not couple to $H$ and $\bar H$
to avoid the tachyonic stop issue mentioned above.
Once this field gets a VEV,
$\left<\hat{S}\right>= m_{S} \ + \ \theta^2 \ F_{S}$, 
superpartner masses can be generated.  Assuming that SUSY breaking is transmitted 
through the mass term for the field used to break $SU(5)$ to the SM, we replace
$
m_{\Sigma} \ \text{Tr} \ \hat{24}_H^2 \  \  \text{by} \  \  \lambda_{\Sigma} \  \hat{S} \ \text{Tr} \ \hat{24}_H^2.
$
In this case SUSY breaking is mediated through $\Sigma_8$ and $\Sigma_3$, but since these
fields do not carry hypercharge, the bino and right-handed charged sleptons, $\tilde{e}^c_i$, remain
massless (at the two-loop level) at the messenger scale while running effects would drive the latter mass to tachyonic values.
A realistic SUSY spectrum then requires transcending the minimal model by introducing extra representations which can be used as messengers.

We proceed by appealing to neutrino masses for guidance.  Explicitly, we compare the possible mechanisms for 
neutrino masses and their role in gauge mediation. Neutrino 
masses can be generated through the Type I~\cite{TypeI-1,TypeI-2,TypeI-3,TypeI-4,TypeI-5}, 
Type II~\cite{TypeII-1,TypeII-2,TypeII-3,TypeII-4,TypeII-5} 
or Type III~\cite{TypeIII,Ma,Borut-Goran-TypeIII,Pavel-III-1,Pavel-III-2,LR-III}
\footnote{For the study of flavour and CP violation in non-renormalizable $SU(5)$ models see for example Ref.~\cite{Borzumati}.} 
seesaw mechanism ($R$-parity violating interactions can also be used to generate neutrino masses but do not provide messenger candidates and we continue by assuming $R$-parity
conservation).  Now, since
the right-handed neutrinos needed for Type I seesaw are SM singlets they cannot generate
a realistic superpartner spectrum, leaving Type II and Type III seesaw as the only viable options. Type II seesaw
necessitates the introduction of two chiral superfields, ${\bf \hat{15}_H} \text{ and }
{\bf\hat{\bar{15}}_H}$, and was shown to produce a realistic spectrum in Ref.~\cite{Rossi}.  On the other
hand Type III seesaw requires introducing only one chiral superfield, a ${\bf \hat{24}}$, making it more minimal
and worthy of study.

Recently, several groups have investigated the Type III seesaw mechanism in the context of $SU(5)$ 
grand unified theories~\cite{Ma,Borut-Goran-TypeIII,Pavel-III-1,Pavel-III-2}. In order to realize this 
mechanism one has to introduce a new matter superfield in the adjoint representation:
\begin{eqnarray}
&& \hat{24}= (\hat{\rho_8},~ \hat{\rho_3},~ \hat{\rho}_{(3,2)},
~\hat{\rho}_{(\bar{3},2)}, ~\hat{\rho_0}) = (8,1,0) \bigoplus(1,3,0) \bigoplus (3,2, -5/6) \bigoplus (\bar{3},2,5/6) \bigoplus(1,1,0).  
\end{eqnarray}
In the context of non-SUSY $SU(5)$, this idea was pursued in a 
non-renormalizable model~\cite{Borut-Goran-TypeIII} and then in a fully renormalizable 
scenario~\cite{Pavel-III-1}. This mechanism was studied for the first time in the context 
of a supersymmetric grand unified theory in Ref.~\cite{Pavel-III-2}.

In our case the relevant superpotential, which is used to generate neutrino masses 
through both the Type III and Type I seesaw mechanisms, is given by
\begin{eqnarray}
\label{W.seesaw}
&&{\cal W}_{4}= h_1 \ \hat{\bar{5}} \ \hat{24} \  \hat{5}_H \ + \  
\hat{\bar{5}} \left(  h_2 \ \hat{24} \ \hat{24}_H  \ + \  
h_3 \  \hat{24}_H \  \hat{24} 
\ + \ h_4 \ {\rm Tr} \left(  \hat{24} \  \hat{24}_H \right) \right) \hat{5}_H \ / \ M_{Pl}.
\end{eqnarray}
The mass of the seesaw fields, $\rho_0$ and $\rho_3$ responsible for Type I and Type III seesaw
respectively, are computed using the following superpotential
\begin{eqnarray}
\label{W24}
&&{\cal W}_{5}= m \ {\rm Tr} \ \hat{24}^2 \ + \ \lambda \ {\rm Tr} \ \hat{24}^2 \hat{24}_H  \ + \
{\cal O}( 24_H / M_{Pl}). 
\end{eqnarray}
Once $SU(5)$ is broken, $\left< 24_H\right> = \textrm{diag} (2,2,2,-3,-3) \  v_{\Sigma} / \sqrt{30}$, 
the above superpotential can be used to compute the masses for the fields in the ${\bf 24}$ multiplet:
\begin{eqnarray}
\label{Mr0}
&& M_{\rho_0}= m \ - \  A ,
\\
&& M_{\rho_3}= m \ - \  3 \  A,
\\
&&M_{\rho_8}= m \ + \  2  \  A,
\\
\label{Mr32}
&& M_{\rho_{(3,2)}}= \ m \ - \  \frac{1}{2} \ A.
\end{eqnarray}
Where $A=\lambda \ v_{\Sigma} / \sqrt{30}$ and we neglect the effect of higher-dimensional operators for simplicity.  The neutrino mass matrix is given by
\begin{equation}
M_\nu^{ij} = \frac{c_i \ c_j}{M_{\rho_0}} \ + \ \frac{b_i \ b_j}{M_{\rho_3}}~,
\end{equation}
where $c_i$ and $b_i$ are a linear combination of the couplings $h_1-h_4$ in Eq.~(\ref{W.seesaw})
\begin{eqnarray}
c^i &=& \frac{v_u}{2} \left(  \frac{3}{2} \ \frac{h_1^i}{\sqrt{15}}  \ - \  \frac{3\sqrt{2}\  v_{\Sigma}}{5 \ M_{Pl}}  
\left( h_2^i \ + \  h_3^i \right)
\ - \  \frac{\sqrt{2} \ v_{\Sigma}}{M_{Pl}} \ h_4^i \right), \\
b^i&=& \frac{v_u}{2} \left(  h_1^i \ - \  \left(  h_2^i \ + \ h_3^i \right)  \frac{3 \sqrt{2} \ v_{\Sigma}}{2 \sqrt{15} M_{Pl}} \right),
\end{eqnarray}
and typically, it is assumed that  $\rho_3$ and $\rho_0$ are around the ``seesaw scale", $M_{\rho_0},
M_{\rho_3} \approx 10^{11-14}$ GeV. Notice that in this case the neutrino masses are generated through the Type I and Type III 
seesaw mechanisms and one neutrino remains massless. The spectrum can 
then have a Normal Hierarchy with $m_1=0$ or an Inverted Hierarchy with $m_3=0$.

In order to transmit SUSY breaking in this scenario 
one needs to replace the mass term for the ${\bf \hat {24}}$ field in Eq.~(\ref{W24}) :
\begin{equation}
    m \ \text{Tr} \ \hat{24}^2 \rightarrow  \lambda_{S}  \  \hat S \ \text{Tr} \ \hat{24}^2~,
\end{equation}
so that both the fermion and scalar components of ${\bf \hat{24}}$ get a squared mass contribution of
$\lambda_S^2 \ m_S^2$
but the scalars get a further mass squared mixing term, $\lambda_S \ F_S \ \tilde{24} \ \tilde{24}$.  The
upshot of this is that a SUSY breaking mass difference exists between the scalars and fermions of the ${\bf \hat{24}}$:
\begin{equation}
    \left| m_{24}^2 - m_{\tilde {24}}^2 \right| = \left| \lambda_S \ F_S \right|~.
\end{equation}
This difference is communicated to the visible sector by the messengers through the mass parameter
$\Lambda \equiv \lambda F_s/M_{\text{Mess}}$.

We again stress that in this scenario, \textit{``$\nu$--GMSB"}, by adding
only one extra chiral superfield, ${\bf \hat{24}}$, we are able to generate neutrino masses in
agreement with experiments and have a consistent mechanism for gauge mediation
since the components of the ${\bf \hat{24}}$ have color, weak and hypercharge charges.
Since Type I seesaw cannot generate masses for the superpartners and 
Type II seesaw needs two chiral superfields \footnote{Notice that the authors in Ref.~\cite{Rossi} 
have more representations since they need $\hat{15}_H$ and $\hat{\overline{15}}_H$, and 
in general their superpotential contains the following terms:
\begin{eqnarray}
W_4^{II} &=& Y_\nu \  \hat{\overline{5}}^T \ \hat{15}_H  \  \hat{\overline{5}} \ + \  \eta  \ \hat{X} \ \text{Tr} \  \hat{\overline{15}}_H  \ \hat{15}_H
\ + \  \mu_1 \  \hat{\overline{5}}^T_H \ \hat{15}_H  \  \hat{\overline{5}}_H \ + \  \mu_2 \ \hat{5}_H^T  \   \hat{\overline{15}}_H  
\  \hat{5}_H \nonumber \\
& + & \lambda_\nu \  \hat{\overline{5}}^T  \   \frac{\hat{24}_H}{M_{Pl}} \  \hat{15}_H \   \hat{\overline{5}} 
\ + \   \lambda_1 \   \hat{\overline{5}}^T_H  \   \frac{\hat{24}_H}{M_{Pl}} \  \hat{15}_H \   \hat{\overline{5}}_H 
\ + \   \lambda_2 \  \hat{{5}}^T_H  \  \frac{\hat{24}_H^T}{M_{Pl}} \   \hat{\overline{15}}_H \  \hat{{5}}_H,  \\
W_{5}^{II} &=&  \lambda_3 \ \text{Tr} \  \hat{\overline{15}}_H \  \hat{24}_H \ \hat{15}_H  
\ +  \   {\cal O} \left( \hat{24}_H/M_{Pl} \right).
\end{eqnarray}  
Then, one could say that they have less parameters only when some of the interactions 
above, which in general are relevant, are neglected. For example, the term $\text{Tr} \  \hat{\overline{15}}_H \ \hat{24}_H \ \hat{15}_H $ 
gives a mass splitting between the messengers after $SU(5)$ is broken.}, 
\textit{Type III seesaw provides the simplest
framework for gauge mediation in $SU(5)$ grand unified theory via seesaw fields}. It is
important to emphasize the differences between this scenario and that studied in
Ref.~\cite{Mohapatra}, where the authors: studied a more involved 
case with several copies of the $24$ field; neglected the very 
relevant interaction---Tr $\hat{24}^2 \hat{24}_H$, which tells us that 
the seesaw scale is large; did not discuss that neutrino masses are 
generated through both the Type I and Type III seesaw mechanisms; 
and did not consider radiative $B$-term generation. In our opinion, 
these are crucial features of our scenario which deserve attention
and we investigate their effects in detail.

\section{$\nu$-GMSB Predictions}
In gauge mediation scenarios it is typically assumed that the messengers are degenerate and
therefore all associated with the same contributions, $\Lambda$, to the soft masses.
Being in a specific GUT model allows us the advantage of
seeing that this not necessarily true.  Here the messengers, the ${\bf \hat{24}}$, 
attain mass splittings from Eq.~(\ref{W24}) due their couplings to the $SU(5)$ breaking ${\bf \hat{24}_H}$.  
The masses are given in Eqs.~(\ref{Mr0}-\ref{Mr32}), where we take
$m\rightarrow \lambda_S \ m_S$ to allow for gauge mediation.  Since these masses differ from
each other, each messenger field 
will have a different $\Lambda$ parameter associated with 
it: $\Lambda_i$ where $i = \left(\rho_0, \rho_3, \rho_8, \rho_{\left(3,2\right)}\right)$. 
In this section, we will for convenience reparameterize the mass relations in terms of
$M_{\rho_3}$ and $\hat{m} = M_{\rho_8} \  /  \ M_{\rho_3}$.  Also, 
any phase in Eqs.~(\ref{Mr0}-\ref{Mr32}) can always be rotated away to yield positive
values for each of the masses.  Then assuming no relevant phase between $m$ and $A$ leads to three
possible cases for this reparameterization:
\begin{itemize}

\item Case I: \ $m \ < \ A/2$;  \ $M_{\rho_{(3,2)}}= \frac{1}{2} M_{\rho_3} \left(1- \hat m \right)$ where $0 < \hat{m} < 1$,

\item Case II: \ $A/2 \ < \ m \ < \ 3 A$;  \ $M_{\rho_{(3,2)}}= \frac{1}{2} M_{\rho_3} \left(\hat m - 1\right)$ where $\hat{m} > 1$,

\item Case III: \  $m \ > \ 3 A$;  \ $M_{\rho_{(3,2)}}= \frac{1}{2} M_{\rho_3} \left(\hat m + 1 \right)$ where $\hat{m} > 0$.

\end{itemize}
For the remainder of the paper we will focus on case III since the range of $\hat m$ is the union of cases I and II.  Specifically we will consider $0.1 < \hat m< 10$ to reduce the
fine tuning between the components of the ${\bf \hat{24}}$.

In general, at each seesaw field threshold, the gaugino masses will receive a one-loop contribution,
which must be evolved down to the next threshold, 
modified by the new contribution and evolved again.  However, the effect from running
between these thresholds is small since the messengers are never separated by 
more than an order of magnitude.  Therefore, one can simply state the gaugino
masses as a boundary condition at $M_\text{Mess} \equiv M_{\rho_3}$.  Computing the gaugino
masses at one-loop yields the following results at the messenger scale:
\begin{eqnarray}
&& M_3  (M_\text{Mess})= a_3 \left(3 \Lambda_{\rho_8} + 2 \Lambda_{\rho_{(3,2)}}\right),
\\
&& M_2  (M_\text{Mess})= a_2 \left(2 \Lambda_{\rho_3} + 3 \Lambda_{\rho_{(3,2)}}\right),
\\ 
&& M_1  (M_\text{Mess})= a_1 \ 5 \ \Lambda_{\rho_{(3,2)}},
\label{eq:gaugino}
\end{eqnarray}
where $a_i=\alpha_i/4\pi$ and $\Lambda_{i} = \lambda_S \ F_S \  /  \ M_{i}$.

Scalar masses are generated at two-loops and can be calculated using the same
philosophy discussed for the gauginos.  In general, scalar masses
will also receive contributions from their Yukawa couplings to the messengers.  However, once
these become sizable, at higher messenger scales, they lead to low energy lepton number violation.
We postpone a study of this effect to a future paper and continue with the assumption that 
$M_{\text{Mess}} \ll 10^{14-15}$  GeV. We will use $M_{\text{Mess}} = 10^{11}$  GeV to illustrate the numerical results.  
This also means that as in minimal models of GMSB, the trilinear $a$-terms and 
bilinear $B$-term will be zero at the messenger scale but non-zero at the SUSY 
scale due to RGE effects. As a result, the boundary conditions for the scalar parameters are:
\begin{eqnarray}
m_{\tilde Q}^2(M_\text{Mess}) &=&
          8 \ a_3^2 \ \Lambda_{\rho_8}^2
        + 3 \ a_2^2 \ \Lambda_{\rho_3}^2
        + \left(\frac{16}{3} \ a_3^2 + \frac{9}{2} \ a_2^2 +\frac{1}{6} \ a_1^2
        \right) \Lambda_{\rho_{(3,2)}}^2,
    \\
m_{\tilde{u}^c}^2(M_\text{Mess}) & = &
          8 \ a_3^2 \ \Lambda_{\rho_8}^2
        + \left(\frac{16}{3} \ a_3^2 + \frac{8}{3} \ a_1^2
        \right) \Lambda_{\rho_{(3,2)}}^2,
    \\
m_{\tilde{d}^c}^2(M_\text{Mess}) & =&
          8 \ a_3^2 \ \Lambda_{\rho_8}^2
        + \left(\frac{16}{3} \ a_3^2 + \frac{2}{3} \ a_1^2
        \right) \Lambda_{\rho_{(3,2)}}^2,
\\
m_{\tilde L}^2(M_\text{Mess}) & = & m_{H_u}^2 =  m_{H_d}^2 =
          3 \ a_2^2 \ \Lambda_{\rho_3}^2
        + \left(\frac{9}{2} \ a_2^2 + \frac{3}{2} \ a_1^2
        \right) \Lambda_{\rho_{(3,2)}}^2,
     \\
m_{\tilde{e}^c}^2(M_\text{Mess}) & = &
          6 \ a_1^2 \ \Lambda_{\rho_{(3,2)}}^2,
     \\
a_i(M_\text{Mess}) & = & 0; \text{    i = u,d,e},
    \\
B(M_\text{Mess}) & =& 0.
\label{eq:squarks}
\end{eqnarray}
See the Appendix for our notation. It is well-known that in gauge mediation,
the masses of all the generations of a given sfermion type are degenerate since
they have the same charges,  \textit{i.e.} $m_{\tilde{Q}_1} = m_{\tilde{Q}_2} = m_{\tilde{Q}_3}$,
while Yukawa effects in the running will push the third generation mass below the degenerate masses
of the first and second generation.  The right-handed components have different masses 
than the left-handed ones because of their different charges.  

Armed with this information we are ready to study the predictions of this model,
focusing on case III.  Calculations are done by inputting the gauge couplings and fermion masses 
at the $Z$ mass scale with a guess for $\tan \beta$ and then evolving up to the messenger scale
using one-loop renormalization group equations (RGEs).  At that scale, the boundary conditions for
the soft terms are calculated and those values are then evolved to the SUSY scale using one-loop
RGEs.  The electroweak symmetry breaking (EWSB) constraints are then used to solve for
$\mu$ and $B$.  If the $B$ value from the EWSB conditions does not match the one given from the RGEs,
a new guess for $\tan \beta$ is used and the process repeats until it converges on a value of $\tan \beta$.

It is important to keep in mind that there are only four input parameters: 
\begin{center}
$\hat{m}$, \  \ $\Lambda \equiv \Lambda_{\rho_3}$, \  \ $M_\text{Mess} \equiv M_{\rho_3}$ \ and \ $\text{sign}(\mu)$,  
\end{center}
so that validation of this scenario could possibly begin once three superpartner masses 
are known and the rest of the spectrum can be calculated.

\subsection{Doubly Radiative Electroweak Symmetry Breaking}
Before diving into the spectrum, it would be useful to contemplate EWSB and the $\mu/B\mu$ problem.
The latter arises in GMSB when generation of the $\mu$ term is linked to SUSY breaking, which usually results in the untenable situation $B \gg \mu$.  However, in our approach, $\mu$ is a parameter of the superpotential that arises from doublet-triplet splitting and we do not attempt to link it to SUSY breaking.

It is common in the literature to assume a value for $\tan \beta$ and then use the EWSB equations to solve for $B$, hence indicating an ignorance of the mechanism which generates this term.  We see no reason to adopt this approach since $B$ is radiatively generated in these SUSY breaking scenarios (we
refer to this as doubly radiative EWSB).  Therefore, even though $B$ is very close to zero at the messenger scale, an appropriate value is generated by RGE running from the messenger to the SUSY scale.  The EWSB equations can then be used to determine the appropriate value of $\tan \beta$ and $\mu$:
\begin{eqnarray}
\label{EWSB.B}
B \ \mu & = &
	\frac{\tan \beta}{1 \ + \ \tan^2 \beta} \ 
	\left(
		2 |\mu|^2
		\ + \ m_{H_u}^2
		\ + \ m_{H_d}^2
	\right),
\end{eqnarray}
\begin{eqnarray}
\label{EWSB.mu}
|\mu|^2 & =&
	- \ \frac{1}{2} M_Z^2
	 \ + \  \frac
	 {
	 	m_{H_u}^2 \ \tan^2 \beta \  -  \  m_{H_d}^2
	}
	{
		1 - \tan^2 \beta
	}.
\end{eqnarray}

Satisfying these equations automatically allows for a nontrivial vacuum and guarantees that the potential is bounded from below:
\begin{align}
	(B \mu)^2 \ &  >  \  \left(  m_{H_d}^2 \ + \  |\mu|^2 \right)  \left( m_{H_u}^2 \ + \  |\mu|^2 \right),
\\
	2 \ B \mu \ & < m_{H_u}^2 \ + \ m_{H_d}^2 \ + \ 2  |\mu|^2,
\end{align}
respectively.

\begin{figure}[ht]
\begin{center}
\includegraphics[width=12.0cm]{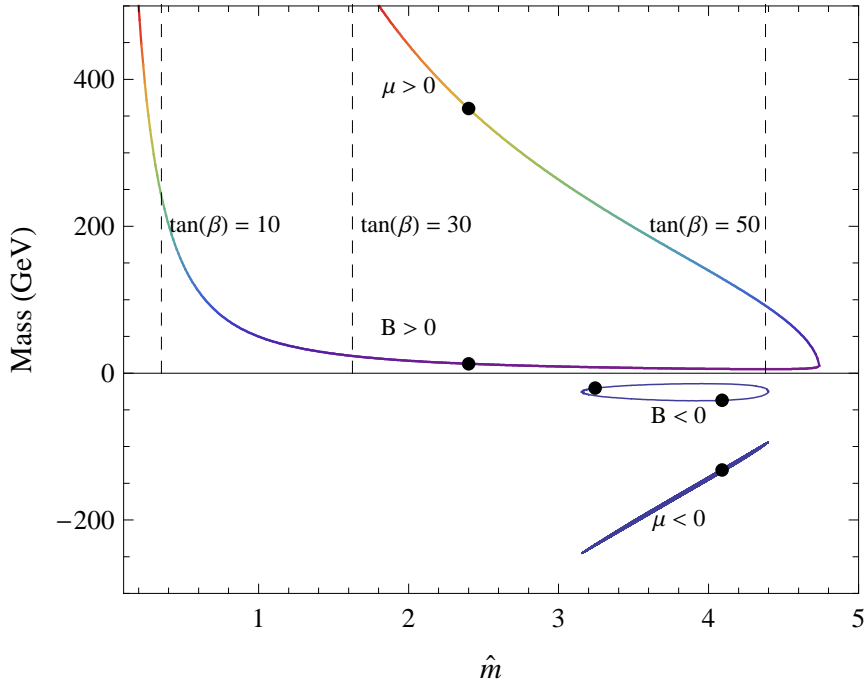}
\caption{
Values of $B$ and $\mu$ at the SUSY scale versus $\hat{m}$ for $\Lambda= 50$ TeV 
and  $M_{\rm Mess} = 10^{11}$~GeV.  Note that $\mu$ and B are both positive (negative)
above the $\hat m$ axis (below the $\hat m$ axis).
The dashed lines represent lines of constant $\tan \beta$
above the $\hat m$ axis only.  Below this axis, one can compare with Fig.~\ref{fig:tanbetaVm};
the top (bottom) of the $B<0$ ellipsoid shape corresponds to
$\tan \beta \sim 38$ ($\tan \beta \sim 22$).  The $\mu < 0$
curve also forms an ellipsoid but a much thinner one, which appears as a line on this plot.
Solutions to the right of the dots on the $\mu > 0$ and $B > 0$ curves are ruled out 
by the constraint on the stau mass, while the same is true of solutions above the two
dots on $\mu<0$ and $B<0$ curves. 
}
\label{fig:BVm}
\end{center}
\end{figure}

\begin{figure}[h!]
\begin{center}
\includegraphics[width=12.0cm]{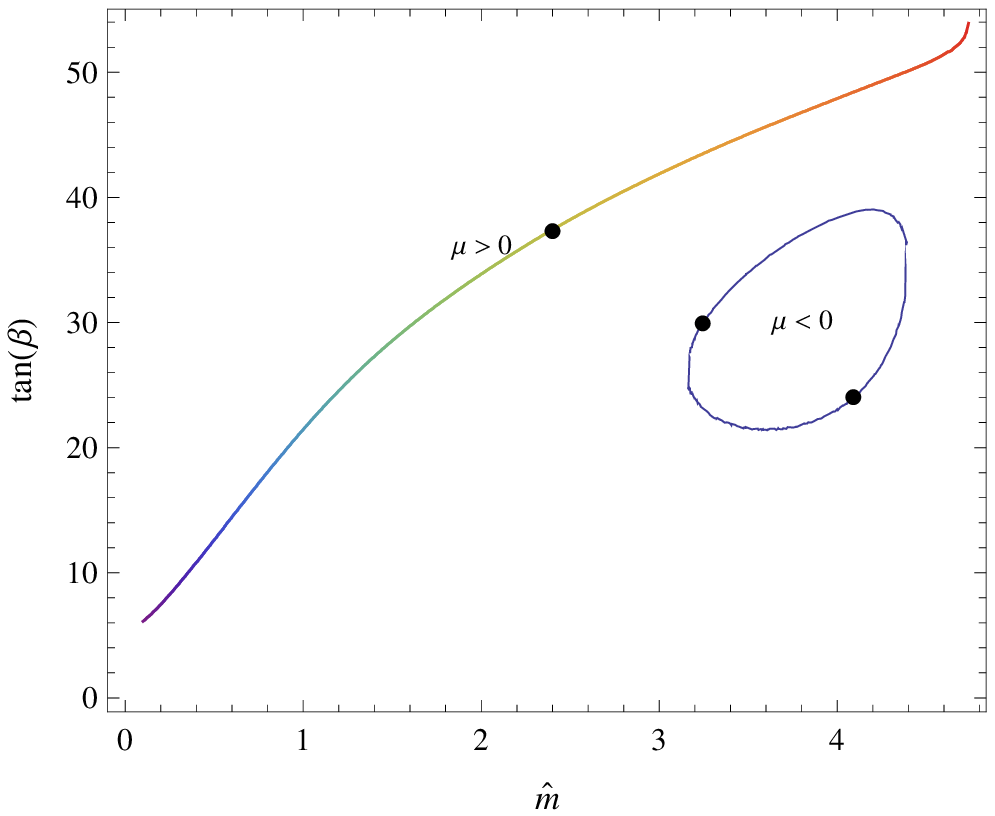}
\caption
{
	Values of $\tan \beta$ versus $\hat m$ for $\Lambda = 50$ TeV and $M_{\rm Mess} = 10^{11}$ GeV.
	The two curves represent, $\mu>0$ and $\mu<0$.  The latter is unique in GMSB with doubly radiative 	EWSB
	and exists in a small part of the parameter space. When $\mu>0$ the solutions above the dot are 
	ruled out by the lower bound on the stau mass. In the case $\mu<0$ one finds consistent solutions
	below the two points in the curve.
}
\label{fig:tanbetaVm}
\end{center}
\end{figure}

In typical models of GMSB, where $B(M_\text{Mess}) = 0$, such as in Ref.~\cite{Babu}, the value of $\tan \beta$ turns out 
to be large\footnote{For a recent phenomenological analysis in this type of scenarios see Ref.~\cite{Abel}. }.  
This is because $\tan \beta$ is inversely related to $B$, which does not run very large.  This is not necessarily the case here as can be seen in Fig.~\ref{fig:BVm}, which plots
$B$ and $\mu$ as a function of $\hat m$ for $\Lambda = 50$ TeV and $M_\text{Mess} = 10^{11}$ GeV.  Solid dashed lines in the upper part of the figure indicates values of constant $\tan \beta$. Solutions to the right
(above) of the dots on the $\mu > 0$ and $B > 0$ ($\mu < 0$ and $B < 0$) curves  are ruled out 
by the constraint on the stau mass.

The behavior of $\tan \beta$ is displayed in Fig.~\ref{fig:tanbetaVm} for the same values of the input parameters.  
The wide range of possible $\tan \beta$ values is due to the $\hat m$ parameter, which reflects the hierarchy 
between the colored and non-colored superpartners: as $\hat m$ increases, this hierarchy decreases. Typical gauge 
mediation models have the same $\Lambda$ for both the colored and non-colored sectors and so correspond 
to $\hat m = 1$ or $\tan \beta \sim 20$  (from Fig.~\ref{fig:tanbetaVm}).  To understand the wider
range of $\tan \beta$ values here it is useful to investigate the largest contributions to the $B$-term beta function:
\begin{equation}
	\label{B.beta}
	\beta_{B} \sim 6 y_t a_t + 6 y_b a_b + \frac{6}{5} g_1^2 M_1 + 6 g_2^2 M_2,
\end{equation}
where the $a$-terms run negative tending to cancel the effects of the gaugino masses, thereby prohibiting $B$ from running too large.  However, since the $a$-terms are mostly driven by the gluino mass parameter, decreasing $\hat m$ increases the gluino mass and allows the $a$-terms to dominate over the electroweak gaugino masses leading to larger positive $B$ values.  This in turn allows for smaller $\tan \beta$ values for small $\hat m$ (LEP2 experiments constrain $\tan \beta > 2.4$~\cite{LEP2}).

Increasing $\hat m$ allows for two options.  The first, if $\mu >0$, which also implies $B > 0$, is that
$\tan \beta$ continues to increase with $\hat m$, as one would naively expect.  The second
is for $\mu, \ B< 0$.  It is due to gaugino masses dominating in $\beta_B$ thereby running
$B$ negative but only when $\tan \beta \lesssim 40$, since larger values would increase $a_b$
in magnitude and allow the $a$-terms to dominate once more.  The region $\mu <0$ is unusual for models of gauge mediation where the $B$-term is generated radiatively.  It is interesting to note that the region of $\mu < 0$ is allowed 
only for a small part of the parameter space: $3.2 \lesssim \hat m \lesssim 4.4$ and $22 \lesssim \tan \beta \lesssim 38$.
For the sake of brevity, we will focus most of the remaining paper on the $\mu > 0$ region
noting here that the major difference between these two regions is the value of $\tan \beta$
which will result in heavier masses for the lightest stau and sbottom in the $\mu < 0$ region.

As a final note, notice that EWSB solutions exist only when $\hat{m} \leq 4.8$ indicating a deep relationship between high scale physics---the mass splittings in the ${\bf \hat{24}}$, $\hat m$--- and the low scale physics---EWSB.

\subsection{Superpartner Spectrum}
As in any gauge mediation mechanism the lightest supersymmetric particle (LSP) 
is the gravitino since its mass is given by, $m_{3/2} \approx F/M_{Pl}\approx 10^2-10^4$ keV, 
where in our case if we take as messenger scale, $M_\text{Mess} \approx 10^{11}$ GeV, 
$\sqrt{F} \approx 10^7-10^8$ GeV is the SUSY breaking 
scale and $M_{Pl} \sim 10^{18}$ GeV, is the reduced Planck scale. 
The rest of the spectrum has some distinctive features from the typical gauge mediation due to $\hat m$.

\begin{figure}[h]
\begin{center}
\includegraphics[width=12.0cm]{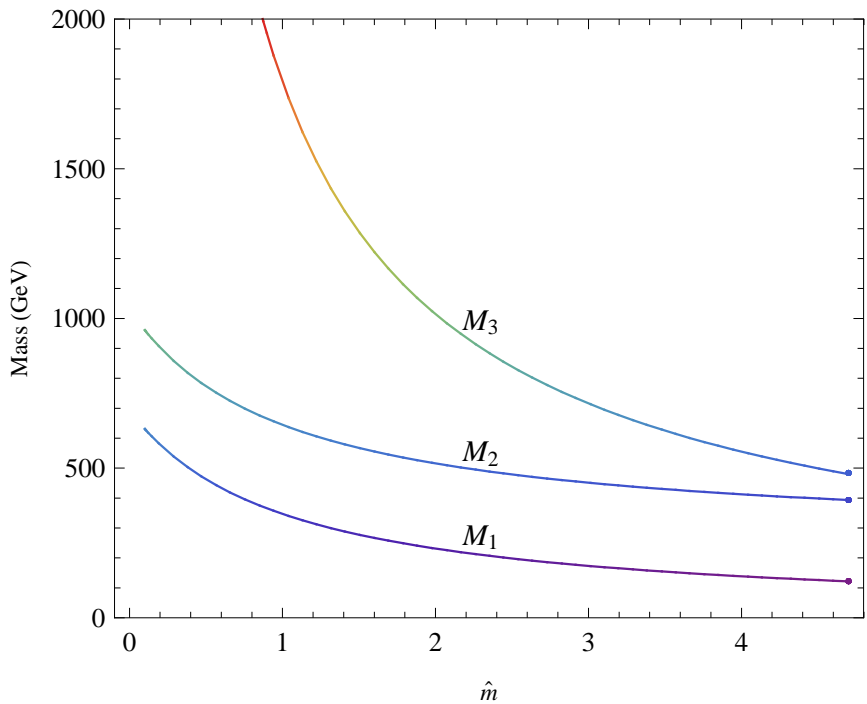}
\caption{
Gaugino mass parameters at the SUSY scale versus $\hat m$ for $\Lambda= 50$ TeV and  $M_{\rm Mess} = 10^{11}$~GeV.}
\label{fig:mgaugino2}
\end{center}
\end{figure}

We begin by examining the gaugino mass parameters versus $\hat m$ at the SUSY
scale for $\Lambda=50$ TeV and $M_{\rm Mess} = 10^{11}$~GeV, Fig.~\ref{fig:mgaugino2}.
This plot reflects the fact that as $\hat m$ increases, the hierarchy
between the colored and non-colored sectors decreases thus reducing the ratio $M_3:M_2:M_1$ from
$20:2:1$ at $\hat m = 0.1$ to $4:3:1$ at $\hat m = 4.5$.

\begin{figure}[h]
\begin{center}
\includegraphics[width=12.0cm]{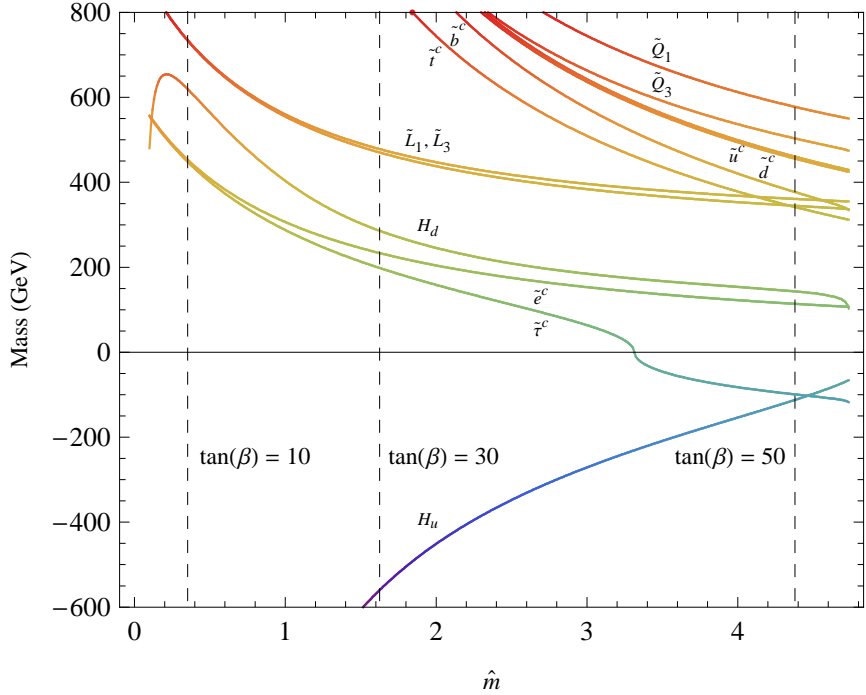}
\caption
{
	Sfemion mass parameters versus $\hat m$ for $\Lambda = 50$ TeV, $M_{\rm Mess} = 10^{11}$ GeV and 
	$\mu > 0$.  The actual values plotted are $\text{sign}(m_\phi^2)\sqrt{|m_\phi^2|}$ so that negative
	values indicate negative mass squared values.  The dashed lines correspond to constant values of  $\tan \beta$.
}
\label{fig:msfermion}
\end{center}
\end{figure}

This effect of $\hat m$ is also reflected in Fig.~\ref{fig:msfermion} where we see the squark
masses drawing closer to the slepton masses as $\hat m$ increases.  We see two other
interesting features as $\hat m$ increases: $m_{H_u}^2$ becomes less negative so that eventually
EWSB would not be possible (as was seen in Fig.~\ref{fig:BVm}) and that the stau mass parameter
eventually becomes tachyonic.  To understand the former behavior, examine the largest contributions
to the $m_{H_u}^2$ beta function:
\begin{equation}
\beta_{m_{H_u}^2} \sim 6 |y_t|^2 (m_{H_u}^2 + m_{\tilde Q_3}^2 + m_{{\tilde t}^c}^2
)
 - 6 g_2^2 |M_2|^2  -  \frac{6}{5} g_1^2 |M_1|^2.
\end{equation}
Typically, $m_{H_u}^2$ runs negative due to the product of the large top Yukawa
coupling with the stop masses.  However, as $\hat m$ increases, this product decreases compared 
to the gaugino masses, eventually $m_{H_u}^2$ does not run negative enough thus spoiling
radiative EWSB.
In fact Fig.~\ref{fig:msfermion} cuts-off when EWSB is no longer possible, at around $\hat m = 4.8$ for $\Lambda = 50$~TeV.  Again, this result is interesting because it specifies that there cannot
be too much splitting in the ${\bf \hat{24}}$ multiplet due to the constraints of EWSB.
The right-handed stau parameter becomes tachyonic for large $\hat m$ because of the large
$\tan \beta$ values, which increase the value of $y_\tau$ running $m_\tau^2$ negative.  This latter feature is more constraining on the parameter space and places the upper bound
$\hat m \lesssim 2.4$ (for stau masses consistent with LEP 2 bounds, $m_{\tilde \tau} > 100$ GeV).

A lower bound on $\hat m$ can also be derived if one wishes to limit the amount of fine tuning
necessary to satisfy EWSB, Eq.~(\ref{EWSB.mu}).  This can be most clearly seen by examining 
this equation in the limit $\tan \beta \gg 1$
\begin{equation}
	\label{EWSB.mu.2}
	|\mu|^2 = -\frac{1}{2} M_Z^2 - m_{H_u}^2.
\end{equation}
The amount of cancellation needed between $\mu^2$ and $m_{H_u}^2$ to produce $\frac{1}{2} M_Z^2$
is a measure of the necessary fine-tuning and increases with the magnitude of $|m_{H_u}^2|$ and
decreasing $\hat m$.  In the interest of fine tuning, we restrict
$|m_{H_u}| \sim |\mu| < 500$ GeV which when combined with the stau bounds lead us to study the range:
\begin{equation}
	1.8 < \hat m < 2.4,
\end{equation}

\begin{figure}[h!]
\begin{center}
\includegraphics[width=12.0cm]{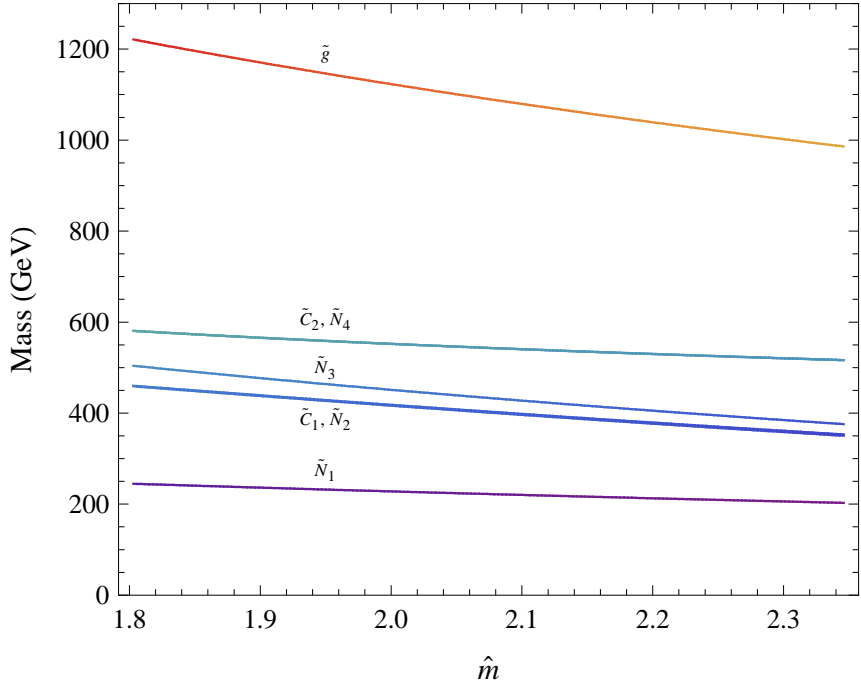}
\caption{Physical neutralino, chargino and gluino masses versus $\hat m$ for $\Lambda = 50$ TeV, $M_{\rm Mess} = 10^{11}$ GeV and $\mu > 0$.  In this plot we focus on the region of reduced fine-tuning,
$1.8 \le \hat m \le 2.4$.}
\label{fig:physics.bosinos}
\end{center}
\end{figure}

The physical spectrum for the gauginos plotted versus $\hat m$ is shown in Fig.~\ref{fig:physics.bosinos}.  
To understand the composition of the neutralinos and charginos first consider
Eq.~(\ref{EWSB.mu.2}) which further reduces to $\mu = |m_{H_u}|$ for $|m_{H_u}|^2 \gg M_Z^2$, typically a good assumption.  Since the Higgsino masses are proportional to $\mu$,
they are also proportional to $|m_{H_u}|$.  Consulting with Figs.~\ref{fig:mgaugino2} and \ref{fig:msfermion} indicates that the neutralinos, from lightest to heaviest are mostly: bino, wino-Higgsino mix, Higgsino and wino-Higgsino mix while the charginos are both wino-Higgsino mixes.  The gluino is the heaviest gaugino for the the value of $\hat m$ shown.

\begin{figure}[h]
\begin{center}
\includegraphics[width=12.0cm]{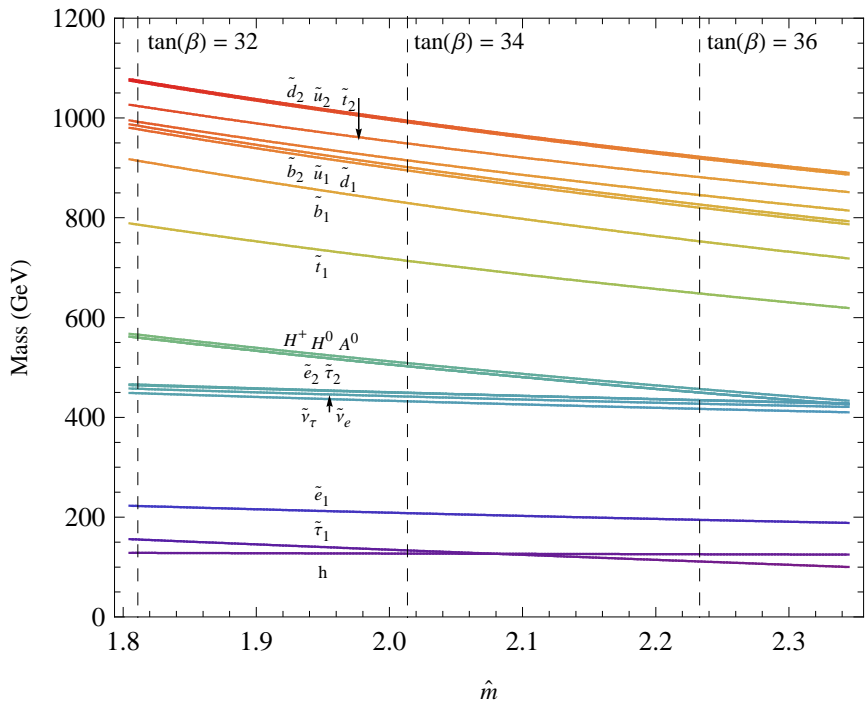}
\caption
{
	Physical sfermion masses versus $\hat m$ for $\Lambda = 50$ TeV, $M_{\rm Mess} = 10^{11}$
	GeV and $\mu > 0$ with dashed lines of constant of $\tan \beta$.
	In this plot we focus on the region of reduced fine-tuning, $1.8 \le \hat m \le 2.4$..
}
\label{fig:physics.sfermions}
\end{center}
\end{figure}

The physical sfermion spectrum is shown in Fig.~\ref{fig:physics.sfermions} with dashed lines of constant $\tan \beta$.  In this region of minimal fine-tuning, the mass ratio of squarks to left-handed sleptons--- $m_{\tilde q}:m_{\tilde l_2}\sim  2:1$.   Furthermore, the Higgs mass is above the LEP 2 lower bound of $114.4$ GeV for this range of $\hat m$ and the most serious constraint comes from the mass of the lightest stau.

Because gaugino masses go as the dynkin index of the messengers while the sfermion masses are
proportional to the square root of the dynkin index, large messengers representations or many copies
of messengers lead to gaugino masses larger than the corresponding sfermion masses.  This applies in
our case where the gluino is heavier than the squarks, the wino heavier than the sleptons and the lightest
stau is the next to lightest supersymmetric particle (NLSP).  The fact that $\tan
\beta$ is large at large $\hat m$ is a further contribution making the stau the NLSP.  Since the coupling of
TeV particles to the LSP gravitino is highly suppressed, the NLSP plays an
important role in collider physics.  Depending on the lifetime of the stau NLSP, it will produce charged
tracks or displayed vertices, both of which would be spectacular signals at the Large Hadron Collider (LHC)~\cite{Manuel}.

\begin{figure}[h]
\begin{center}
\includegraphics[width=12.0cm]{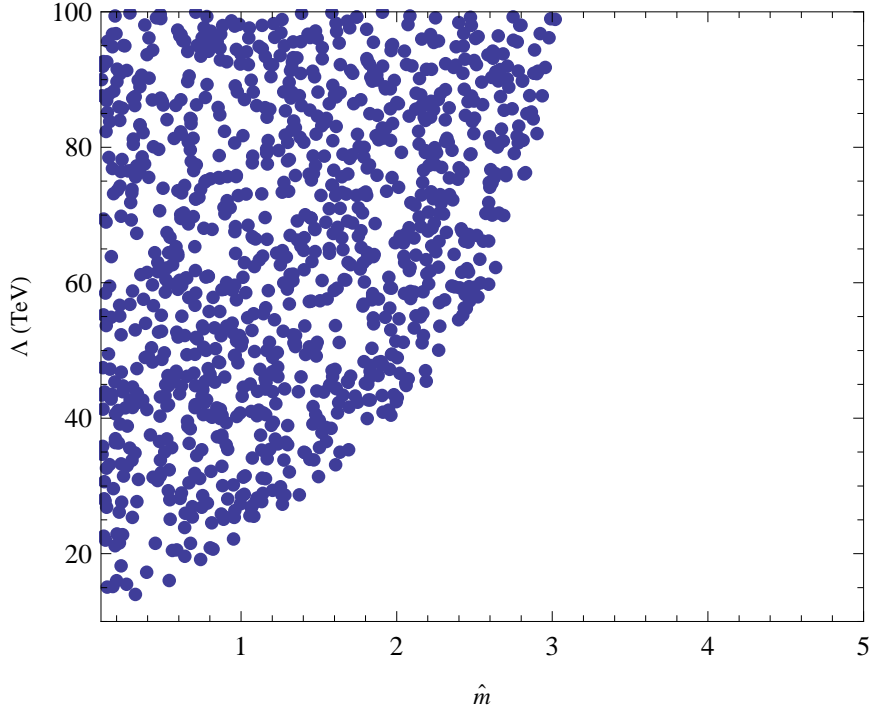}
\caption
{The dots indicate the allowed parameter space in the $\hat m - \Lambda$ plane given collider constraints on the supersymmetric 
masses and EWSB for  $\mu > 0$ and $M_\text{Mess} = 10^{11}$ GeV.}
\label{Parameter-space}
\end{center}
\end{figure}

As a final attempt to familiarize the reader with the features of the spectrum, we list the
masses for  $\hat{m}=2$ in Table \ref{table.mspectrum}, which reflects the features noted thus far.  We
also give a similar table for $\mu < 0$ and $\hat m = 4$, Table \ref{table.mspectrum.negativemu} to get a
feeling for this part of the parameter space.  Since this region has smaller values of $\tan \beta$, the
bounds on the lightest stau are satisfied even with $\hat m$ larger than the range discussed above and
also leads to a less hierarchical spectrum. We also note that our results for $\tan \beta$ are consistent with the constraints coming from $Y_b=Y_\tau$ unification. See for example Ref.~\cite{Ross}.

Finally, in order to understand the predictions in the full parameter space 
we show the allowed range in the $\hat m - \Lambda$ 
plane given collider constraints on the supersymmetric masses 
and EWSB for  $\mu > 0$ and $M_\text{Mess} = 10^{11}$ GeV in Fig.~\ref{Parameter-space}.
The stau is the NLSP in the entire parameter space.

\begin{table}[h]
\caption{
\label{table.mspectrum.Half}
Sparticle mass spectrum at the SUSY scale for $\tan \beta = 34, \ \Lambda = 50 \text{ TeV},
\ M_{\rm Mess} = 10^{11} \text{ GeV}$, $\hat m = 2$ and $\mu >0$.  First and second generation masses are degenerate.
}
\begin{tabular}{|l|l|l|}
\hline
 Particle     & Symbol   & Mass (GeV)         \\
\hline \hline
stop     &$\stilde{t}_1$,~~ $\stilde{t}_2$          &718,~~ 953 \\
sbottom     &$\stilde{b}_1$,~~ $\stilde{b}_2$          &834, ~~920  \\
up squarks     &$\stilde{u}_1$,~~ $\stilde{u}_2$          &906, ~~996 \\
down squarks     &$\stilde{d}_1$,~~ $\stilde{d}_2$          &900, ~~999 \\
\hline \hline
stau, ~~tau sneutrino   &$\stilde{\tau}_1$,~~ $\stilde{\tau}_2$,~~
$\stilde{\nu}_{\tau}$     &135, ~~450, ~~433 \\
selectron,~~electron sneutrino   &$\stilde{e}_1$,~~ $\stilde{e}_2$,
~~$\stilde{\nu}_e$           &209, ~~450, ~~443 \\
\hline \hline
neutralinos   &$\stilde{N}_1$,~~ $\stilde{N}_2$,~~
$\stilde{N}_3$,~~ $\stilde{N}_4$~~       &228, ~~418, ~~451, ~~552 \\
charginos  &$\stilde{C}_1$,~~ $\stilde{C}_2$    &416, ~~552 \\
\hline \hline
gluino    &$\tilde{g}$       &1123                  \\
\hline \hline Higgses    &$m_{A^0}$, ~~$m_{H^{\pm}}$, ~~$m_{H^0}$, ~~$m_{h^0}$
&506, ~~512, ~~506, ~~127 \\ \hline
\end{tabular}
\label{table.mspectrum}
\end{table}

\begin{table}[h]
\caption{
\label{table.mspectrum.Half}
Sparticle mass spectrum at the SUSY scale for $\tan \beta = 22, \ \Lambda = 50 \text{ TeV},
\ M_{\rm Mess} = 10^{11} \text{ GeV}$, $\hat m = 4$ and $\mu <0$.  First and second
generation masses are degenerate.
}
\begin{tabular}{|l|l|l|}
\hline
 Particle     & Symbol   & Mass (GeV)         \\
\hline \hline
stop     &$\stilde{t}_1$,~~ $\stilde{t}_2$          &368,~~ 612 \\
sbottom     &$\stilde{b}_1$,~~ $\stilde{b}_2$          &478, ~~564  \\
up squarks     &$\stilde{u}_1$,~~ $\stilde{u}_2$          &498, ~~610 \\
down squarks     &$\stilde{d}_1$,~~ $\stilde{d}_2$          &494, ~~615 \\
\hline \hline
stau, ~~tau sneutrino   &$\stilde{\tau}_1$,~~ $\stilde{\tau}_2$,~~
$\stilde{\nu}_{\tau}$     &106, ~~369, ~~359 \\
selectron,~~electron sneutrino   &$\stilde{e}_1$,~~ $\stilde{e}_2$,
~~$\stilde{\nu}_e$           &130, ~~371, ~~363 \\
\hline \hline
neutralinos   &$\stilde{N}_1$,~~ $\stilde{N}_2$,~~
$\stilde{N}_3$,~~ $\stilde{N}_4$~~       &111, ~~155, ~~167, ~~429 \\
charginos  &$\stilde{C}_1$,~~ $\stilde{C}_2$    &141, ~~429 \\
\hline \hline
gluino    &$\tilde{g}$       &629                  \\
\hline \hline Higgses    &$m_{A^0}$, ~~$m_{H^{\pm}}$, ~~$m_{H^0}$, ~~$m_{h^0}$
&357, ~~366, ~~357, ~~118 \\ \hline
\end{tabular}
\label{table.mspectrum.negativemu}
\end{table}

\section{Constraints from gauge coupling unification and proton decay}

\begin{center}
\begin{figure}[th]
\begin{center}
\includegraphics[width=8.0cm]{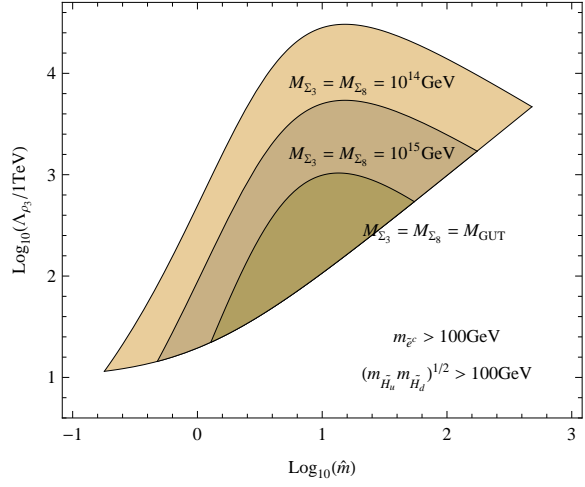}
\caption{Parameter $\Lambda_{\rho_3}$ as a function of the mass
splitting $\hat m$, and different values of the masses of the $\Sigma_3$
and $\Sigma_8$ fields.}
\label{fig:Lambda3}
\end{center}
\end{figure}
\begin{figure}[h]
\begin{center}
\includegraphics[height=6.7cm]{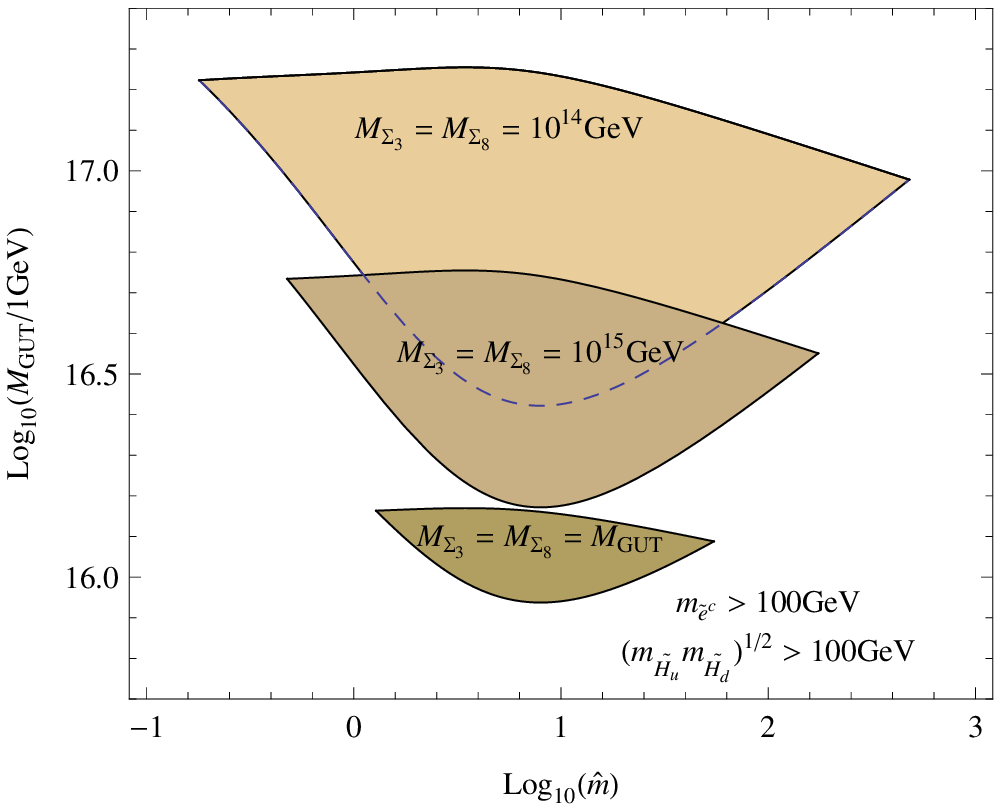}
\caption{Unification scale as a function of the mass splitting $\hat{m}$,
and different values of the masses of the $\Sigma_3$ and $\Sigma_8$ fields.}
\label{fig:mgut_mess}
\end{center}
\end{figure}
\end{center}

We study in this section the possible constraints obtained by requiring 
unification of the gauge couplings when the gaugino and 
squark masses are determined by gauge mediated SUSY breaking mechanism proposed in this paper.
For the general constraints in minimal SUSY $SU(5)$ see Ref.~\cite{German}.
Let us analyze the case where $M_T=M_V=M_{\rm GUT}$, and 
leave $M_{\Sigma_3}$ and $M_{\Sigma_8}$ as free parameters. 
Solving the RGE's in Eqs.(\ref{eq:alpha1gut})-(\ref{eq:alpha3gut}), 
we find
\begin{widetext}
\begin{eqnarray}
M_{\rm GUT} &=& M_Z \left[ \hat{m}^{-6} \left( \frac{1+\hat m}{2}\right)^{12}
\frac{M_Z^{20} \, m_{\tilde{e}^c}^3 \, m_{\tilde{u}^c}^3}
{m_{\tilde Q}^6 \, M_{\tilde W}^4 \, M_{\tilde g}^4 
\, M_{\Sigma_3}^6 \, M_{\Sigma_8}^6}
\exp\left[2\pi
\left( 5 \alpha_1^{-1}-3 \alpha_2^{-1} - 2 \alpha_3^{-1}\right) (M_Z) \right]
\right]^{1/24}~.
\end{eqnarray}
\end{widetext}
Notice that the unification scale does not depend explicitly on the absolute value
of the masses of the $\rho$ multiplet, but on their mass splitting 
$\hat m = M_{\rho_8}/M_{\rho_3}$. Remember that if $\hat{m} > 1$, the $\rho_3$ 
field is the lightest partner of the $\hat{24}$ representation, and 
$\rho_8$ is the heaviest, while the opposite is true when 
$\hat{m} < 1$. The corresponding gauge coupling at the unification 
scale is given by
\begin{widetext}
\begin{eqnarray}
\alpha_{\rm GUT}^{-1}(M_{\rm GUT}) &=& \frac{1}{24} \Bigg\{
\left( -25 \alpha_1^{-1} + 15 \alpha_2^{-1}
+ 34 \alpha_3^{-1}\right)(M_Z) \nn \\ &&
+ \frac{1}{\pi} \log
\left[{\hat m}^{51}\left(\frac{1+\hat m}{2}\right)^{-6}
\frac{M_{\rho_3}^{60} \, m_{\tilde Q}^{27} \, m_{\tilde d^c}^{6} \, 
M_{\tilde W}^{10} \, M_{\tilde g}^{34} \, M_{\Sigma_3}^{15}
\, M_{\Sigma_8}^{51} }
{M_Z^{194} \, m_{\tilde u^c}^{3/2} \, m_{\tilde e^c}^{15/2}} \right] \Bigg\}~,
\end{eqnarray}
\end{widetext}
and contrary to the unification scale it depends on the absolute 
value of the masses of the $\rho$ multiplet. The unification scale and 
the gauge coupling at the unification scale are both independent of the 
Higgsino masses $m_{\tilde H_u}$ and $m_{\tilde H_d}$. 
The product of their masses is, however, constrained by unification:
\begin{widetext}
\begin{equation}
m_{\tilde H_u} \, m_{\tilde H_d} = M_Z^2 \left[ 
\hat{m}^{54} \left( \frac{1+\hat m}{2}\right)^{12}
\frac{M_Z^{28} \, m_{\tilde{e}^c}^9 \, m_{\tilde{u}^c}^{21} 
\, m_{\tilde{d}^c}^{12} \, M_{\tilde g}^{36} \, M_{\Sigma_8}^{54}}
{m_{\tilde Q}^{30} \, m_{\tilde L}^{12} \, M_{\tilde W}^{44} 
\, M_{H_u}^4 \, M_{H_d}^4 \, M_{\Sigma_3}^{66}}
\exp\left[6\pi
\left( 5 \alpha_1^{-1}-11 \alpha_2^{-1} +6  \alpha_3^{-1}\right) (M_Z) \right]
\right]^{1/8}~.
\end{equation}
\end{widetext}
By imposing a lower bound on the product of the Higgsino masses,
the latter condition sets, as a function of the $\rho$ mass splitting 
$\hat m$, an upper limit on the parameter $\Lambda_{\rho_3}$.
Furthermore, since the lightest sfermion at the messenger scale 
is ${\tilde e}^c$, and by using \Eq{eq:squarks}, a lower 
bound on $\Lambda_{\rho_3}$ can be obtained from 
a given value of $m_{{\tilde e}^c}$, neglecting the 
running of its mass .

In Fig.~\ref{fig:Lambda3} we show the allowed values of
$\Lambda_{\rho_3}$ which are compatible with the limits
$m_{{\tilde e}^c} \ > \ 100$~GeV,
with $M_{\rm Mess}=10^{11}$~GeV, and $(m_{{\tilde H}_u} m_{{\tilde H}_d})^{1/2} > 100$~GeV. 
Under these conditions, and for $M_{\Sigma_3}=M_{\Sigma_8}=M_{\rm GUT}$, 
the parameter $\Lambda_{\rho_3}$ is constrained 
and can take values only in the range
\begin{equation}
25~{\rm TeV} < \Lambda_{\rho_3} < 1000~{\rm TeV}~,
\end{equation}
which is fairly independent of the messenger scale
because the gauge couplings run rather slowly
at very high energy scales. The corresponding 
$\rho$ mass splitting is constrained to be in the range
\begin{equation}
1.3 < \hat m < 50~.
\end{equation}
These limits, however, can be relaxed if the $\Sigma_3$ and $\Sigma_8$
fields are lighter than the unification scale. Indeed, for 
$M_{\Sigma_3}=M_{\Sigma_8}=M_{\rm GUT}$ the unification 
scale is of the order of $10^{16.1}$~GeV, which might be in conflict 
with proton decay~\cite{review} if we do not suppress the couplings 
of the colored triplets mediating proton decay to matter. The unification scale 
becomes larger, and compatible with proton decay, if these two fields 
become lighter. This is illustrated in Fig.~\ref{fig:mgut_mess}, showing the allowed 
values of the unification scale as a function of $\hat m$, 
assuming $M_{\Sigma_3}=M_{\Sigma_8}$ at or below $M_{\rm GUT}$.
As it has been discussed in detail in Ref.~\cite{review} the lower bound 
on the mass of the colored triplet mediating proton decay is basically 
$M_T > 10^{17}$ GeV  if no additional suppression mechanism 
is used. Notice that this is perhaps the simplest solution to suppress proton decay 
since in this case one does not have mixings between the squarks of the 
different families. However, since in Eq.~(\ref{W1}) we assume the 
existence the higher-dimensional operators, one can always suppress 
the dimension five contributions to proton decay using the fact that 
the couplings of the colored triplets to matter are free parameters in general.
 
\begin{center}
\begin{figure}[th]
\begin{center}
\includegraphics[height=6.7cm]{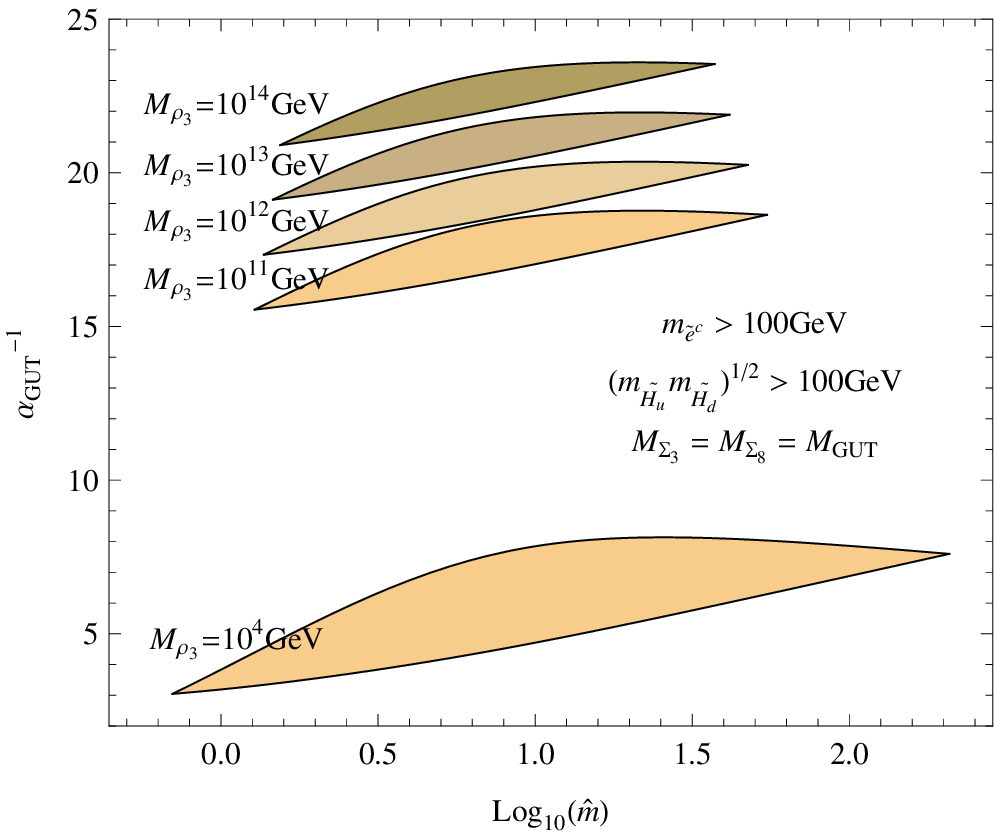}
\put(-30,165){(a)}
\includegraphics[height=6.7cm]{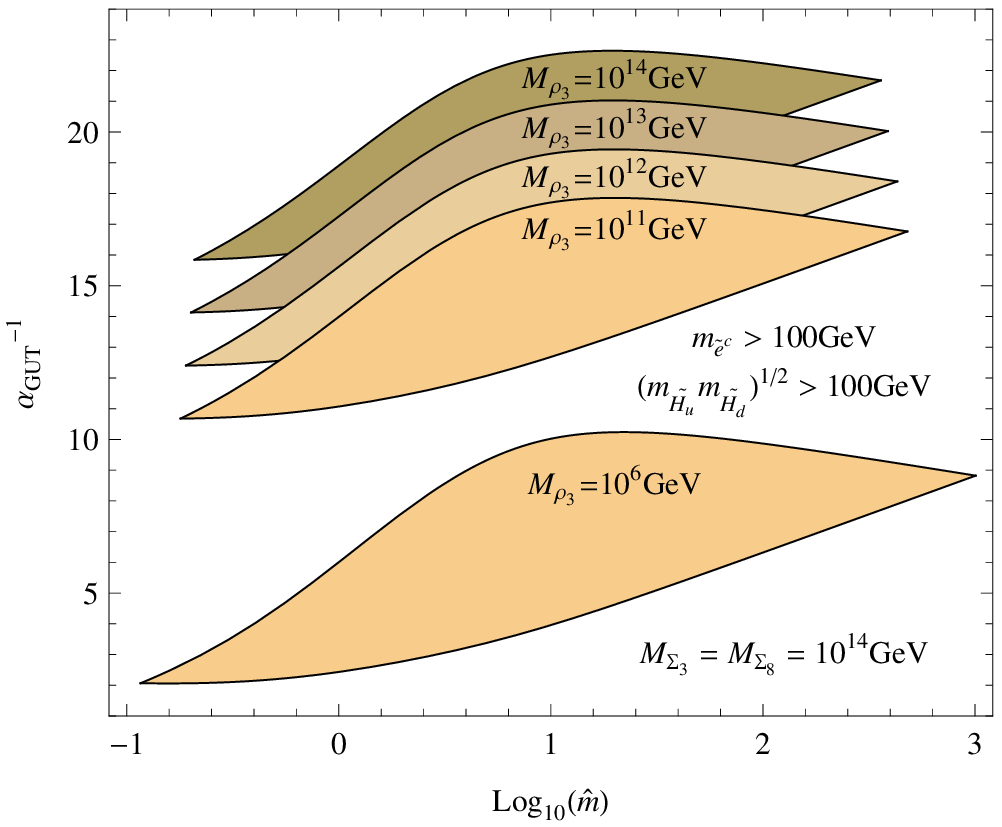}
\put(-30,165){(b)}
\caption{Gauge coupling at the unification scale as a function of the mass splitting $\hat{m}$, and of the mass of the $\rho_3$ field 
for (a) $M_{\Sigma_3}=M_{\Sigma_8}=M_{\rm GUT}$,  and  (b) $M_{\Sigma_3}=M_{\Sigma_8}=10^{14}$~GeV.}
\label{fig:alpha_mess}
\end{center}
\end{figure}
\end{center}

The gauge coupling at the unification scale is represented in 
Figs.~\ref{fig:alpha_mess}(a) and~\ref{fig:alpha_mess}(b) for different 
values of the $\rho_3$ mass, and different choices of $M_{\Sigma_3}$ 
and $M_{\Sigma_8}$. It is worth mentioning that GMSB together with 
unification requires the Higgsino masses and $m_{{\tilde e}^c}$
to be relatively light. 

Here we have seen that suppressing proton decay by pushing up the GUT scale 
requires the $\Sigma_3$ and $\Sigma_8$ fields to be below 
the GUT scale, in particular only when these fields are below $10^{14}$ GeV 
one can achieve unification at $10^{17}$ GeV.  Notice that using these 
results one can find a lower bound on the messenger scale which is $M_{\rho_3} > 10$ TeV, 
see Fig.~\ref{fig:alpha_mess}.
However, such low-scale gauge mediation in this context one requires fine-tuning the
messenger masses because of the Tr $\hat{24}^2 \hat{24}_H$ term, which 
tells us that the masses of the seesaw fields should be very large.  

\section{Summary and Outlook}
We have presented a simple scenario for gauge mediated supersymmetry breaking where 
the messengers are the fields that generate neutrino masses. We refer to this mechanism 
as ``$\nu$-GMSB". In this scenario the neutrino masses are generated through the 
Type I and Type III seesaw mechanisms and in the simplest 
case where the contributions of  Yukawa couplings to soft masses are not considered we find:

\begin{itemize}

\item Sparticle and Higgs masses are predicted from only 
four free parameters: $\hat{m}$ defining the splitting in the ${\bf 24}$ representation,
the messenger scale $M_\text{Mess}$,
$\Lambda=\lambda_S \ F_S / M_\text{Mess}$
and \ $\text{sign}(\mu)$.

\item EWSB is achieved through a doubly radiative mechanism, 
where the $B$-term is very small at the messenger scale  and radiatively generated at the SUSY scale.  
resulting in a constrained spectrum.

\item Imposing ``minimal" fine-tuning, $100 \  \text{GeV} \leq  |\mu|,  |m_{H_u}| (M_Z) \leq 500$ GeV, 
EWSB conditions and collider constraints,  and $M_\text{Mess}\approx 10^{11}$ GeV, we find small mass 
splitting between the messengers and the mostly right-handed stau is always the NLSP.

\item For $\mu < 0$, $\hat m$ and tan $\beta$ are in a small 
range leading to a very constrained spectrum. For example, when $\Lambda=50$ TeV 
and $M_\text{Mess}=10^{11}$ GeV, $3.2 \lesssim \hat m \lesssim 4.4$ 
and $22 \lesssim \tan \beta \lesssim 38$. 

\item LHC signatures include charged tracks as is typical in for GMSB with a stau NLSP.
It is well-known, that this allows for the reconstruction of the gaugino and squark masses to determinate the spectrum.

\item The lower bound on the messenger scale from 
the constraint $\alpha_{\rm GUT} < 1$ is $M_{\rho_3} > 10$ TeV.

\item In a future publication we plan to study in this model
the predictions and/or constrains from rare decays,
the baryogenesis and leptogenesis mechanism, and the analysis
of the Yukawa coupling contributions to the soft masses.
 
\end{itemize}

\begin{widetext}

{\textit{Acknowledgments}}:
{\small We would like to thank B. Bajc, V. Barger, G. Shiu and X. Tata for enjoyable 
discussions and comments. 
The work of P.F.P. was supported in part by the U.S. Department of Energy 
contract No. DE-FG02-08ER41531 and in part by the Wisconsin Alumni 
Research Foundation. 
The work of H. I. was supported in part by the Scientific Research Foundation for the Returned Overseas Chinese Scholars, State Education Ministry.
G.R. acknowledges support by the Ministerio de Ciencia e
Innovaci\'on under Grant No. FPA2007-60323, and CPAN
(Grant No. CSD2007-00042), by the Generalitat
Valenciana under Grant No. PROMETEO/2008/069, and
by the European Commission MRTN FLAVIAnet under
Contract No. MRTN-CT-2006-035482.
S.S. is supported in part by the U.S. Department 
of Energy under grant No. DE-FG02-95ER40896,
and by the Wisconsin Alumni Research Foundation.}
\appendix
\section*{APPENDIX: NOTATION AND RGES}
\label{ap:gauge}
In order to set our notation we include the superpotential of the MSSM:
\begin{equation}
W_{\rm MSSM} = \hat{u}^c  y_u {\hat Q} {\hat H_u} + \hat{d}^c y_d
{\hat Q} {\hat H}_d + {\hat e}^c y_e {\hat L} {\hat H}_d + \mu {\hat
H}_u {\hat H}_d \> , \label{MSSMsuperpot}
\end{equation}
where $ y_{u,d,e}$ are matrices in family space, and the soft 
SUSY-breaking Lagrangian is given by
\begin{eqnarray}
\lagr_{\rm soft}^{\rm MSSM} &=& - \frac{1}{2} \left ( M_3 \ \tilde
g \ \tilde g  \ + \ M_2 \ \tilde W \  \tilde W  \ + \ M_1 \  \tilde B \  \tilde B + \textrm{h.c.}
\right )
\nonumber \\
& - & \left( \tilde{u}^c \ a_u  \  \tilde{Q} \ H_u + \tilde{d}^c \  a_d
\  \tilde{Q} \  H_d  \ +  \  \tilde{e}^c  \ a_e \  \tilde{L} \ H_d
\ + \ B \mu \ H_u \ H_d  \ + \  \textrm{h.c.} \right)
\nonumber \\
& - &  \tilde{Q}^{\dagger} \ m_{\tilde Q}^2 \  \tilde{Q} \ - \ 
\tilde{L}^{\dagger} \ m_{\tilde L}^2 \  \tilde{L} \ - \  \tilde{u}^{c
\dagger} \ m_{{\tilde u}^c}^2 ~ \tilde{u}^c  \ - \ \tilde{d}^{c \dagger} \
m_{{\tilde d}^c}^2 \ \tilde{d}^c \ - \ \tilde{e}^{c \dagger} \ m_{{\tilde
e}^c}^2 \  \tilde{e}^c
\nonumber \\
& - &  m_{H_u}^2  \ H_u^{\dagger} \ H_u \ - \ m_{H_d}^2 \ H_d^{\dagger} \ H_d ~,
\label{MSSMsoft}
\end{eqnarray}
where $M_3$, $M_2$, and $M_1$ are the gluino, wino, and bino mass
terms. The second line in Eq. (\ref{MSSMsoft}) contains the $(\rm
scalar)^3$ couplings, and $a_{u,d,e}$ are complex $3 \times 3$
matrices in family space. Here again $m_{\tilde Q}^2$, $ m_{\tilde
L}^2$, $m_{{\tilde u}^c}^2$, $ m_{{\tilde d}^c}^2$, and $m_{{\tilde
e}^c}^2$ are $3 \times 3$ matrices in family space.

The renormalization group of equations (RGEs) for the gauge couplings 
in this model are given by
\begin{eqnarray}
\alpha_1^{-1} (M_Z) & = &  \alpha_{\rm GUT}^{-1} + \frac{1}{2\pi}
\left(
    4  \ln \frac{M_{\rm GUT}}{M_Z}
+  \frac{3}{10} \ln \frac{M_{\rm GUT}}{m_{\tilde L}} +  \frac{3}{5}
\ln \frac{M_{\rm GUT}}{m_{{\tilde e}^c}} + \frac{1}{10} \ln
\frac{M_{\rm GUT}}{m_{\tilde Q}} +  \frac{4}{5} \ln \frac{M_{\rm
GUT}}{m_{{\tilde u}^c}}  \right.
\nonumber \\
&& \hspace*{20mm} \left.    +   \frac{1}{5} \ln \frac{M_{\rm
GUT}}{m_{{\tilde d}^c}} + \frac{1}{10} \ln \frac{M_{\rm
GUT}}{m_{H_d}} + \frac{1}{10} \ln \frac{M_{\rm GUT}}{m_{H_u}}
 + \frac{1}{5} \ln \frac{M_{\rm
GUT}}{m_{{\tilde H}_d}} \right.
\nonumber \\
&& \hspace*{20mm} \left.  + \frac{1}{5} \ln \frac{M_{\rm
GUT}}{m_{{\tilde H}_u}}  - 10 \ln \frac{M_{\rm GUT}}{M_V} +
\frac{2}{5} \ln{\frac{M_{\rm GUT}}{M_T}} + 5 \ln \frac{M_{\rm
GUT}}{M_{\rho_{(3,2)}}} \right)~, \label{eq:alpha1gut}
\end{eqnarray}
\begin{eqnarray}
\alpha_2^{-1} (M_Z) & = &  \alpha_{\rm GUT}^{-1} + \frac{1}{2\pi}
\left( - \frac{20}{6}  \ln \frac{M_{\rm GUT}}{M_Z} +  \frac{1}{2}
\ln \frac{M_{\rm GUT}}{m_{\tilde L}} +  \frac{3}{2} \ln \frac{M_{\rm
GUT}}{m_{\tilde Q}}  +   \frac{1}{6} \ln \frac{M_{\rm GUT}}{m_{H_d}}
+   \frac{1}{6} \ln \frac{M_{\rm GUT}}{m_{H_u}}
 \right.
\nonumber \\
&& \hspace*{20mm} \left. +  \frac{1}{3} \ln \frac{M_{\rm
GUT}}{m_{{\tilde H}_d}}  +  \frac{1}{3} \ln \frac{M_{\rm
GUT}}{m_{{\tilde H}_u}} +  \frac{4}{3} \ln \frac{M_{\rm
GUT}}{M_{\tilde W}} - 6 \ln \frac{M_{\rm GUT}}{M_V} \right.
\nonumber \\
&& \hspace*{20mm} \left. + 2 \ln \frac{M_{\rm GUT}}{M_{\Sigma_3}}
  + 2 \ln \frac{M_{\rm GUT}}{M_{\rho_3}} +
3\ln \frac{M_{\rm GUT}}{M_{\rho_{(3,2)}}} \right)~,
\label{eq:alpha2gut}
\end{eqnarray}
\begin{eqnarray}
\alpha_3^{-1} (M_Z) &=& \alpha_{\rm GUT}^{-1} + \frac{1}{2\pi}
\left( - 7 \ln \frac{M_{\rm GUT}}{M_Z} +  \ln \frac{M_{\rm
GUT}}{m_{\tilde Q}} + \frac{1}{2} \ln \frac{M_{\rm GUT}}{m_{{\tilde
u}^c}} + \frac{1}{2} \ln \frac{M_{\rm GUT}}{m_{{\tilde d}^c}}
\right.
\nonumber \\
&& \hspace*{20mm} \left. + 2 \ln \frac{M_{\rm GUT}}{M_{\tilde g}} -
4 \ln \frac{M_{\rm GUT}}{M_V} + \ln{\frac{M_{\rm GUT}}{M_T}} +  3
\ln \frac{M_{\rm GUT}}{M_{\Sigma_8}}  \right.
\nonumber \\
&& \hspace*{20mm} \left.  + 2 \ln \frac{M_{\rm
GUT}}{M_{\rho_{(3,2)}}}
 + 3 \ln \frac{M_{\rm GUT}}{M_{\rho_8}} \right)~. \nonumber \\
\label{eq:alpha3gut}
\end{eqnarray}
\end{widetext}

\end{document}